\documentclass[review]{elsarticle}

\usepackage{graphicx}			% allows us to import images
\usepackage{amssymb,amsmath,amsthm,bm} 
\usepackage{bbm}
\usepackage{hyperref}
\usepackage{graphicx} % Allows including images
\usepackage{booktabs} % Allows the use of \toprule, \midrule and \bottomrule in tables
\usepackage{graphicx}
\usepackage{lineno,hyperref}
\usepackage{graphicx}			% allows us to import images
\usepackage{amssymb,amsmath,amsthm,bm} 
\usepackage{hyperref}
\usepackage{graphicx} % Allows including images
\usepackage{booktabs} % Allows the use of \toprule, \midrule and \bottomrule in tables
\usepackage{graphicx}
\usepackage{url}
\usepackage{adjustbox}

\numberwithin{equation}{section}
\usepackage{natbib}
\usepackage{float}
\usepackage{textcmds}
\usepackage{threeparttable}
\usepackage[T1]{fontenc}
\usepackage{lipsum}

\theoremstyle{definition}

\usepackage{booktabs}
\usepackage[table,xcdraw]{xcolor}
\usepackage{multirow}
\usepackage{rotating,tabularx}
\usepackage[bottom]{footmisc}
\journal{Social Networks}

\biboptions{authoryear}

\usepackage{color}

\setcitestyle{authoryear,open={(},close={)}}

\begin{document}

\begin{frontmatter}

\title{Selection of Exponential-Family Random Graph Models via Held-Out Predictive Evaluation (HOPE)}

 \author[add1]{Fan Yin}
   \ead{yinf2@uci.edu}
      \author[add2]{Nolan Edward Phillips}
   \ead{nolanephillips@fas.harvard.edu}
   \author[add1,add3]{Carter T. Butts\corref{cor1}}
   \ead{buttsc@uci.edu}
% Please address correspondence to 
   \cortext[cor1]{Please address correspondence to Carter T. Butts, 2145 Social Science Plaza A,
 University of California, Irvine, CA 92697, USA}
      \address[add1]{Department of Statistics, University of California, Irvine, CA 92697, USA}
   \address[add2]{Department of Sociology, Harvard
           University, Cambridge, MA 02138, USA}
   \address[add3]{Departments of Sociology, Computer Science, and EECS, and Institute for Mathematical Behavioral Sciences, University of California, Irvine, CA 92697, USA}

\begin{abstract}

Statistical models for networks with complex dependencies pose particular challenges for model selection and evaluation. In particular, many well-established statistical tools for selecting between models assume conditional independence of observations and/or conventional asymptotics, and their theoretical foundations are not always applicable in a network modeling context.  While simulation-based approaches to model adequacy assessment are now widely used, there remains a need for procedures that quantify a model's performance in a manner suitable for selecting among competing models.  Here, we propose to address this issue by developing a predictive evaluation strategy for exponential family random graph models that is analogous to cross-validation.  Our approach builds on the held-out predictive evaluation (HOPE) scheme introduced by \citet{wang2016multiple} to assess imputation performance.  We systematically hold out parts of the observed network to: evaluate how well the model is able to predict the held-out data; identify where the model performs poorly based on which data are held-out, indicating e.g. potential weaknesses; and calculate general summaries of predictive performance that can be used for model selection. As such, HOPE can assist researchers in improving models by indicating where a model performs poorly, and by quantitatively comparing predictive performance across competing models. The proposed method is applied to model selection problem of two well-known data sets, and the results are compared to those obtained via nominal AIC and BIC scores. 
\end{abstract}


\end{frontmatter}

% \linenumbers
\nolinenumbers

\section{Introduction}
The exponential family random graph modeling (ERGM) framework \citep{holland1981exponential, frank1986markov, snijders2006new, hunter2006inference} (known in older work by the term $p^{*}$ \citep{wasserman1996logit}) has emerged as an important approach to the statistical analysis of social network data, providing a highly general way of specifying distributions on graphs and allowing the complex dependence structure of edges in a network to be specified in terms of local structural properties \citep{robins2007recent}. A wide variety of features have been proposed as potential instantiations of the different types of driving forces governing the formation of social networks \citep{morris2008specification}, with the potential to accommodate an increasingly rich range of network types.  At the same time, however, poor specifications can lead to unrealistic model behavior \citep{handcock2003assessing, schweinberger2011instability},  and in practice considerable domain expertise can be required to select terms that implement the correct dependence structure for a specific setting \citep{lusher2013exponential}.  

Choosing among competing model specifications can be viewed as a model selection problem, making information criteria such as Akaike Information Criterion (AIC) \citep{akaike1973information} and Bayesian Information Criterion (BIC)\citep{schwarz1978estimating} natural choices for adjudication. However, these criteria rely on a number of theoretical assumptions that are frequently problematic in a network modeling context.  First, edge variables in typical network models are non-independent, making it difficult to determine the effective sample size needed for size-corrected AIC and BIC calculations \citep{hunter2008goodness}; indeed, at this time the theoretical justification for these criteria is unclear in the case of models for single networks with dyadic dependence \citep[though see][for some possible directions]{kolaczyk2015question,schweinberger2017exponential}. Second, likelihood calculations for complex ERGMs rely on stochastic approximations (e.g., bridge sampling) that become expensive for large networks and where high precision is needed.  This is not a barrier to parameter estimation (which typically relies on quantities such as likelihood ratios between identically specified models with similar parameters that can be more precisely computed), but makes fine distinctions between the likelihoods of similarly performing but differently specified models difficult.  

The theoretical (if not computational) challenges noted above can be avoided in a Bayesian context by performing model selection by Bayes factors \citep{raftery1995bayesian}.  ERGM applications to date require fairly expensive, high-quality posterior simulation using methods such as the \emph{Reversible Jump Exchange Algorithm} \citep{caimo2013bayesian}, a tailored version of the conventional \emph{Reversible Jump MCMC} algorithm \citep{green1995reversible} combined with the \emph{exchange algorithm} \citep{Murray:2006:MDD:3020419.3020463} by \citet{caimo2011bayesian} to deal with the double intractability of the posterior density, as the ERGM likelihood cannot be computed analytically in general. This approach has the advantage of being theoretically principled in the setting of fixed sample size, but the need to obtain a high-quality approximation of the Bayes factor can be computationally demanding. Promising directions in this area include efforts like those of \citet{bouranis2018bayesian} to reduce the computational cost by adjusting the posterior samples yielded by an analytically tractable approximation of the ERGM likelihood, e.g. the pseudolikelihood. However, its magnitude adjustment step requires approximating the \emph{true} ERGM likelihood evaluated at MLE, which raises the same issue encountered in the calculation of AIC/BIC; thus, an efficient and easily used method for obtaining Bayes factors for general ERGMs remains elusive.  Moreover, the Bayes factor itself is not always an ideal tool for model selection.  The Bayes factor (and its multi-model generalizations) provides an answer to the question, ``which of a set of proposed models is more likely to be the true data generating process?'' assuming that the models being evaluated are \emph{a priori} equally probable and that one is correct.  While the equal probability assumption (i.e., uniform model priors) can be adjusted, the hidden assumption that one proposed model is correct -- or, at least, that among incorrect models the model ``more likely'' to be correct is also ``better'' -- is not entirely innocent \citep{Bernardo94,spiegelhalter2002bayesian}.  When no available model is correct, the model preferred by the Bayes factor may or may not have other desirable properties (e.g., better predictive performance for some task of interest), and indeed the Bayes factor may heavily weight aspects of model performance that are not in practice those most valued by the analyst.  

Relatedly, the Bayes factor can be sensitive to model features, such as the tail weight of the parameter priors, that are often chosen on semi-arbitrary grounds (and that in practice often have little impact on estimation). This creates the risk that model selection will be unduly influenced by choices of the analyst that are difficult to constrain and that are otherwise of minimal substantive importance. Moreover, the approach is only applicable to Bayesian inference, which is not at present widely used for ERGMs due to computational challanges.  Thus, while the Bayes factor can be an important tool in the network analyst's arsenal, it also poses considerable difficulty in practice. 

As an attempt to compensate for the difficulties of likelihood-based criteria, \citet{goodreau2007advances,hunter2008goodness} introduced graph-level ``goodness-of-fit'' (GOF) plots to assess the fit of ERGMs, in which several graph-level statistics (e.g., degree distributions, edgewise shared partner distributions) of observed networks are compared against those of simulated networks from the estimated model. The underlying idea is that draws from a fitted ERGM should have structural properties similar to the observed one, and, in particular, that network properties not explicitly used to fit the model (``out of model'' properties) should be reproduced by those that were used (``in model'' properties).  The properties used to assess a model are usually chosen on substantive grounds, though some efforts have been made to suggest relatively ``generic'' statistics of broad utility \citep[e.g.][]{hunter2008goodness,wang2013exponential1, wang2013exponential2, shore2015spectral}.  While this approach has proven useful in practice, it is properly a \emph{model adequacy checking} strategy rather than a \emph{model selection} strategy: it provides ways to identify performance deficiencies in a chosen model, but it does not provide a general rubric for choosing among competing models.  Likewise, the GOF approach is not designed to provide strong information regarding the predictive performance of a fitted model.  Rather, it only answers the question of how well networks drawn from a model fit to a specific data set reproduce other features of that data set.  This is useful for detecting when a fitted model is \emph{incapable} of producing realistic behavior, but it does not establish that the model will predict well (either in the context of extrapolation to new structures or interpolation of held-out data).

While it could perhaps be argued that predictive performance is not always a major concern for ERGMs, lack of predictive power at the very least suggests limitations of a model that should be borne in mind when using it.  Moreover, predictive performance is clearly a consideration in many applications.  For example, studies of international conflict \citep{hoff2004modeling,maoz2006structural} or bill cosponsorship \citep{fowler2006connecting, cranmer2011inferential} are concerned with consequential relations for which predictions are of significant interest.  These could include e.g. the ability to forecast future network states from past network states (or merely from covariates), or conditional prediction of edge states given covariates and/or information on other edges.  In the latter case, performance involving specific edge states (as opposed e.g. to unlabeled properties such as the degree variance) is of obvious importance: it is important to know whether a predicted conflict is between e.g. China and United States versus Bulgaria and Croatia.  A model that successfully reproduced the degree distribution for a given year's conflict network might not perform well at predicting the degree distribution for the next year's network, nor at predicting who will be in a conflict with whom.  Such a model might be judged satisfactory via conventional adequacy checks, but its value for understanding global conflict would be questionable at best.

The above suggest more explicit predictive metrics as potentially useful tools for model selection (as well as model assessment).  In many fields, cross-validation (CV) techniques have been fruitfully used in this role, allowing one to assess how well the predictions from a model generalize to a new data set; flexible, easily understood, and able to be linked directly with performance outcomes of substantive interest, CV methods are well-adapted to model selection \citep[see][for a recent review]{arlot2010survey}. Typically, CV divides the data set into a \emph{training set} (to which the model is fit) and a \emph{test set} (against which the fitted model's predictions are evaluated under pre-specified loss functions), selecting the model with the smallest estimated loss.  Many variants of this procedure exist (e.g., leave-one-out CV, $k$-fold CV, bootstrap CV, etc.), but all share the common feature of assessing predictive performance on a data subset that was held out during parameter estimation. Classical CV for regression models with independent and identically distributed (i.i.d.) data was proposed as early as \citet{geisser1975predictive}, and CV procedures have been tailored for the purpose of performance evaluation in latent variable modeling of relational data \citep[e.g.,][]{hoff2008modeling, dabbs2016comparison, li2016network, chen2018network}, where edge variables are conditionally independent given latent variables and hence can be straightforwardly held out. Likewise, there is work on applying CV to ERGMs where multiple networks from the same population model are available, and entire networks can be held out \citep{stewart2019multilevel}.  Such work suggests considerable potential utility in applying CV to the more typical setting of general ERGMs on single graphs realizations.

% in a relational context
A difficulty with standard CV techniques in a general ERGM setting is the direct dependence of edge variables, which makes it impossible to simply omit edge variables without changing the underlying probability model.\footnote{This is related to the inconsistency of dependence models under naive subsampling, when the presence of unmeasured vertices is not accounted for \citep{shalizi2013consistency}; procedures that do allow consistent estimation are discussed by \citet{schweinberger2017exponential}.}  Intuitively, the \emph{presence} of an edge variable in such a model is itself informative, and this information must be retained for predictions to be meaningful.  This same issue arises in the context of ERGM estimation from networks with missing edge data, where the presence of edge variables must still be accounted for even when their states are unknown.  \citet{handcock2010modeling} introduced an estimation scheme for handling such data in the case of ignorable missingness, which resolves this difficulty by integrating over the unknown states of the missing edge variables (and thereby preserving the impact of their interactions with the variables whose states are observed).  The missing data case suggests the key to obtaining CV-like procedures for ERGMs: while one cannot meaningfully hold out edge variables, one can hold out the \emph{states} (i.e., whether a given edge variable contains an edge or a null) of edge variables, retaining their presence but treating them as missing data.  Building on this intuition, \citet{wang2016multiple} proposed a held-out evaluation scheme for evaluating model-based imputation that was analogous to CV, which they dubbed Held-Out Predictive Evaluation (HOPE).  Unlike CV, the edge variables in the validation set under HOPE are only marked as missing (i.e. NA) in the model training phase, instead of being completely eliminated.  Thus, the trained model accounts for the presence of the edge variables, but it is not given information on their states. Testing is then performed by conditional prediction of the held-out edge states from the fitted model conditional on the edge variables that are not held-out. Since the core techniques needed to perform this procedure (estimation with missing edge data and conditional procedure) are supported in standard ERGM software, it can be used without the need for custom software implementations or other special considerations.

While HOPE was originally introduced in the context of imputation assessment, it is a general CV-analogue and the idea of holding out a portion of data as missing can be used for a wide range of evaluation tasks. For instance, \citet{koskinen2018outliers} proposed a model-based approach for the identification of influential nodes in a network by assessing the sensitivity of estimated parameters when all edge variables associated with the corresponding node is held out. In this paper, we introduce the use of HOPE to perform model selection for ERGMs, with an emphasis on simple metrics and procedures that are applicable to a wide range of network data. Using HOPE, researchers can gain information on how well the model is able to predict held-out portions of the data from other edge observations; where the model performs poorly based on held-out data, which can point to weaknesses in parameterization; and how one model's predictive performance compares to others. Because such predictive assessments automatically correct for overfitting (which by definition improves in-sample performance while harming out-of-sample performance), they can be quantitatively compared across specifications in a way that other metrics often cannot.  Taken together, these assessments can assist researchers in improving models and facilitate the comparison of fit across multiple models, thus facilitating model selection.

The remainder of the paper is structured as follows. In section 2, we review the ERGM framework with special emphasis on the estimation of ERGM in the presence of missing data, accompanied by some notation necessary for the purpose of the proposed methodology. In section 3, we elaborate on the idea of HOPE and introduce two general strategies for holding-out data along with a set of metrics that measure a model's performance based on several structural features and at different granularities. Section 4 provides two data examples that demonstrate the effectiveness of HOPE and compare it to existing approaches. Finally, we conclude with a discussion and potential directions for future work.  

\section{ERGM estimation and simulation process}
Consider a random graph $G$ of fixed order $n$, where $V = \{1, 2, \cdots, n \}$ is the node (vertex) set and $Y$ is the stochastic \emph{adjacency matrix} of $G$, such that $Y_{i,j}$ is the variable representing the state of an edge variable from node $i$ to node $j$. For valued networks, edge variable $Y_{i,j}$ represents the edges' strengths or frequencies, e.g. trade flows between countries or numbers of interactions between individuals. In most social networks, self-edges are absent or not of substantive interest, and therefore we assume $Y_{i,i} = 0, i = 1, \cdots, n$ in this paper (though ERGMs can still be fit to networks with loops, and HOPE is still applicable). Lower case $y_{i,j}$ will represent the realized value of the $i,j$ edge variable. Letting $\mathcal{Y}_{n}$ be the set of all possible values of such matrices, an ERGM form for the probability mass function of $Y$ is given by

\begin{equation} 
\label{eq:ERGM}
P_{\eta}(Y=y) = \frac{ h(y) \exp( \eta(\theta)^\intercal g(y) ) } { \kappa(\eta(\theta)) },
\end{equation}

where

\begin{equation}
\kappa(\eta(\theta)) = \sum_{y' \in \mathcal{Y}} h(y') \exp( \eta(\theta)^\intercal g(y') ). 
\end{equation}

The user-defined statistic $g : \mathcal{Y}_{n} \rightarrow \mathbb{R}^{p}$ can incorporate any finite network statistics ranging from exogenous covariates (e.g., homophily based on nodes' characteristics) to endogenous network effects (e.g., transitive closure) that are represented as counts of those local features in the observed network. The model parameter vector is $\theta \in \Theta \subseteq \mathbb{R}^{q}$ and its mapping to canonical parameters is $ \eta : \theta \rightarrow \mathbb{R}^{p} $. For notational convenience, we assume that all covariates $X$ are implicitly incorporated into $g$. The reference measure $h$ determines the support and the basic shape of corresponding ERGM distribution \citep[see][for an illustration]{krivitsky2012exponential}. While the above can be applied to a wide range of relational data types, our HOPE procedure is generally applicable to any network model for which the estimation of parameters in the presence of missing data is feasible. To avoid notational and related complications caused by valued edges, however, we illustrate the key idea of HOPE using canonical binary ERGMs with counting measures (i.e., $\eta(\theta) \equiv \theta$ and $h(y) \equiv \mathbbm{1}_{\mathcal{Y}_{n}} (y)$).

While various approaches can be used to estimate $\theta$ given an observation $y^{obs}$ from $Y$, we focus here on the maximum likelihood estimation (MLE, presently the most widely used class of answer).  Importantly, the MLE can be obtained when part of the data (i.e., edge states) are ignorably missing, which is a critical component of HOPE. To estimate $\theta$ where $Y$ is possibly only partially observed, we follow the scheme introduced by \citet{handcock2010modeling} for constructing the \emph{face-value likelihood} based solely on the observed part of the data, 

\begin{equation}\label{eq:ergm_missing}
P_{\theta}(Y^{obs}=y^{obs}) = \frac{\sum_{y^{mis} \in \mathcal{Y}_{n}(y^{obs}) } \exp(\theta^\intercal g(y^{mis} \cup y^{obs} ) ) } {\kappa(\theta)} 
\end{equation}
where $y^{obs}$ is the \qq{observed} portion of $y$, and $\mathcal{Y}_{n}(y^{obs})$ represents all configurations for \qq{missing} portion of the data that are concordant to $y^{obs}$. Note \eqref{eq:ergm_missing} is a valid likelihood as it correctly describes the probability of the observed data marginalizing across the unobserved portion, and we can estimate the model parameters $\theta$ under relatively general conditions \citep[see][for more details]{handcock2010modeling}. 

% This paragraph should explain how one simulate from the fitted model
Maximizing \eqref{eq:ergm_missing} with respect to $\theta$ yields the MLE, $\hat{\theta}$, from which one can conduct conditional ERGM simulations \citep{handcock2010modeling} to predict (or impute) the states of missing edge variables given the states of observed (i.e., retained) edge variables,
\begin{equation}\label{eq:cond_simulation}
    P_{\hat{\theta}}(Y^{mis} = v | Y^{obs} = y^{obs}) = \frac{\exp [ \hat{\theta}^{\intercal} g(v \cup y^{obs}) ]}{\kappa(\hat{\theta} | y^{obs})}, \ \ \ \ \ v \in \mathcal{Y}_{n}(y^{obs})
\end{equation}

where $\kappa(\hat{\theta} | y^{obs}) = \sum_{u \in \mathcal{Y}_{n}(y^{obs}) } \exp [ \hat{\theta}^{\intercal} g(u \cup y^{obs}) ]$. 

The software package MPNet \citep{wang2014mpnet} and the \texttt{ergm} package \citep{hunter2008ergm} of the statnet \citep{handcock2008statnet} software suite for \textbf{R} \citep{R2018} have implemented simulation-based algorithms for approximating the MLE of $\theta$ under model \eqref{eq:ergm_missing}. Additionally, both allow simulations from the estimated parameters with or without missing data (the former is used in goodness-of-fit assessments). 

In this paper, we implement the HOPE procedure in \textbf{R} using the \texttt{ergm} package. To run conditional simulations from ERGMs, the observed part of the graph can be fixed using the \texttt{constraint} argument in the \texttt{simulate.ergm} function. The procedure is repeated for each set of held-out edge variables in observed data, and the simulated networks are then evaluated based on several criteria from various aspects. This makes HOPE analogous to cross-validation in that we are evaluating how well we can predict data that are not used for model training. \newline 

\section{Held-out Predictive Evaluation}
In this section, we first describe two possible ways that relational data can be held-out from an adjacency matrix. We then elaborate on the implementation procedure for HOPE. Finally, we introduce some generic metrics that can be used to evaluate a model's performance at different granularities.\newline

\subsection{Held-out strategies and general procedure}
To perform HOPE for the purpose of model selection, the entire index set of edge variables $\mathcal{D} = \left\{(i,j) | i,j \in V, i<j \right\}$ ($\mathcal{D} = \left\{(i,j) | i,j \in V, i \neq j\right\}$, if $Y$ is directed) is partitioned into $M$ subsets, $A_1, \cdots, A_M$, where $\bigcup_{i=1}^{M} A_i \subseteq \mathcal{D}$. Intuitively, our procedure operates by holding out the values of edge variables in one subset while fitting to the remainder, using the resulting estimate to predict the values of the held-out data.  (Compare with conventional cross-validation, in which the variables themselves - and not merely their values - are held out.  While this distinction is immaterial for independence models, it is consequential for typical ERGMs.)  While other schemes are also feasible, two natural options for holding out network data are random sampling of edges and removal of all edge values associated with a randomly chosen vertex. For convenience of notation, we assume below that network size is an even number; however, generalization to the odd-numbered case is immediate. 

\begin{enumerate}
    \item Random edge removal: $\mathcal{D}$ is randomly divided into $M$ non-overlapping, equally-sized batches of size $k = {n \choose 2} / M$ edge variables (if the graph is undirected), $k = n (n-1) / M$ edge variables are selected (if the graph is directed), with each batch being held-out in turn. A similar strategy was used by \citet{hoff2008modeling} for the purpose of selecting the optimal number of dimensions in latent space structure. It is worth mentioning that leave-one-out can be argued as a special case for random sample where $k=1$, as each edge variable forms a batch by itself and is held out successively. 
    \item Node removal: All edge variables involving a particular node are simultaneously held-out. \citet{koskinen2018outliers} used a similar strategy to identify influential observations (nodes) under the ERGM framework.
\end{enumerate}

The general procedure for HOPE then proceeds as follows. For $m = 1,\cdots,M$:

1. Fit an ERGM to partially observed data $y^{obs}_{A_{m}^c}$ based on \eqref{eq:ergm_missing}, which in turn gives the corresponding held-out data MLE, $\hat{\theta}^{(m)}$.

2. Simulate $B$ graphs from the conditional distribution $P_{\hat{\theta}^{(m)}} ( \cdot | Y_{ A_{m}^c } = y^{obs}_{A_{m}^c} )$ defined in \eqref{eq:cond_simulation} as $\hat{G}^{m,1}, \cdots, \hat{G}^{m,B}$ with their corresponding adjacency matrices $\hat{y}^{m,1}, \cdots, \hat{y}^{m,B}$. Note, that for any $b \in \{ 1,2,\cdots, B \}$, $\hat{y}^{m,b}_{A_{m}^c} =  y^{obs}_{A_{m}^c}$, because all edge variables in $A_{m}^c$ are fixed when simulating from the conditional distribution \eqref{eq:cond_simulation}.

3. Evaluate the ability of the model to accurately predict the held-out data under any of several error metrics. The choice of which metrics to use will depend on which structural features users deem substantively important for their model.  

When the graph is large, a subset of $\mathcal{D}$, namely $\mathcal{D}_{s}$, can be considered to reduce the computational burden, with metrics being calculated only on these subsets.  In such cases, both uniformly random and stratified subset selection approaches may be considered; the former is simpler (and always applicable), but the latter may lead to more efficient approximation of the complete-partition solution.

\subsection{Error metrics}
It is natural to assess the predictive accuracy of a model at three levels: the dyad, the node, and the graph as a whole.  Here, we outline a number of broadly useful choices of error metrics at each level of assessment. 

\subsubsection{Dyad level metrics}
Dyad-level metrics focus on how well the model performs in predicting the presence and absence of edges.  Our suggested measures are based on two types of edge predictions.  First, we may simply consider whether an edge is predicted to be present or absent in any given simulated draw.  Comparing the true states of held-out edges in each iteration (present/absent) with their imputed states leads to an aggregate \emph{confusion matrix,} $F$, a two-by-two table of true versus predicted values over all $m,b$.  While the confusion matrix is pleasingly direct, it does not discriminate between differences in predicted probabilities and may be information-poor in sparse networks.  For that reason, we also employ marginal probability estimates of edge presence, as computed from held-out models.  Specifically, for $\forall (k,l) \in A_{m}$, we estimate the marginal probability $\Pr(Y_{k,l} = 1 | Y_{A_{m}^c} = y^{obs}_{A_{m}^c})$ by $\hat{y}_{k,l} = \frac{0.5 + \sum_{b=1}^{B} \hat{y}^{m,b}_{k,l}}{B+1}$, which is equivalent to the posterior mean for the binomial probability under a Jeffrey's prior; the use of a local Bayesian estimate here provides shrinkage in the case of highly sparse predictions, and ensures that predictive summaries remain well-behaved even for poor models. It is worth mentioning that the marginal probability $\Pr(Y_{k,l} = 1 | Y_{A_{m}^c} = y^{obs}_{A_{m}^c})$ can be calculated exactly via \emph{change scores} \citep{snijders2006new} under leave-1-out, but we do not pursue this direction as experiments show that $\hat{y}_{k,l}$ is usually a good approximation to the marginal probability when $B$ is sufficiently large (e.g. 500). 

From the above, we may then assess predictive performance by applying standard accuracy assessments to $F$ and/or $\hat{y}$.  In this paper, we employ predictive accuracies for edges (Edge ACC), nulls (Null ACC), and edge variables (Overall ACC) based on $F$, as well as the total squared loss (TSL),
$$ \text{TSL}(y^{obs}, \hat{y}) = \sum_{m=1}^{M} \sum_{(k,l)\in A_{m}} ( \hat{y}_{k,l} - y^{obs}_{k,l} )^2  $$
based on $\hat{y}$.  These metrics are widely used for binary data, easily interpreted, and broadly comparable across models.  As discussed below, they constitute a natural starting point for model selection.

\subsection{Node level metrics}
Node level metrics provide information on how well (or poorly) the model can reproduce the roles and positions of the nodes in the observed graph, when each node's edge variables are successively held out. If a model provides a plausible approximation to the mechanism giving rise to the observed network, then one would anticipate that the discrepancy between the node level metrics of the observed and simulated networks would be minimal. However, this will likely depend on the specific node metric, since some are more sensitive to minor changes in the structure than others (e.g.; degree versus betweenness). Because holding out the state of even one edge variable could potentially alter the roles/positions of all the nodes in a graph, the discrepancy is here calculated for all nodes across all simulations. \newline

Given any node-level centrality measure, $C(v): V \rightarrow \mathbb{R}$, it is natural to assess the discrepancy between the vector of observed node level indices $\bm{C}^{obs} = ( C^{obs}(1), C^{obs}(2), \cdots, C^{obs}(n) )$ and those of simulated, $\bm{C}^{t,b}$. We quantify the degree of agreement via the \emph{coefficient of reliability},  

\begin{equation}\label{eq:reliability_NLI}
\rho_{C} = 1 - \frac{MSE_{C}}{\overline{TSS}_{\bm{C}^{obs}}}
\end{equation}

where $TSS_{\bm{C}^{obs}} = \sum_{i=1}^{n} ( C^{obs}(i) - \bar{\bm{C}}^{obs} )^2$ is the total sum of squares with respect to node level index $C$ measuring the overall variation in the observed values and the mean squared error (MSE) of node level index 
$C$ is given by

$$ MSE_{C} = \frac{\sum_{m=1}^{M} 
\sum_{b=1}^{B} \sum_{i=1}^{n} [ C^{m,b}(i) - C^{obs}(i) ]^{2} }{MBn} $$. 
The reliability metric is similar to the coefficient of determination ($R^2$) in multiple regression and the reliability measures widely employed in psychometric contexts \citep[e.g.][]{gulliksen1987theory}. The coefficient of reliability standardizes the amount of predictive error in a model's node level index relative to the overall variation in the observed values; it can thus be thought of as assessing the extent to which the out-of-sample centrality scores are sufficiently accurate to discriminate between central and peripheral vertices in the observed graph. We focus on degree, betweenness, and eigenvector centralities since they are commonly used metrics in analyzing social networks' structures, but other node-level indices \citep[see, e.g.][]{wasserman1994social} are certainly feasible.  In this context, each index may be motivated as follows:

\begin{itemize}

\item \emph{Degree centrality} quantifies the number of ties that each node has to all other nodes within the network. A node with a higher degree centrality has more ties to the other nodes in the network; accuracy with respect to this index thus indicates that the model is able to distinguish more active vertices from those with fewer relationships.

\item \emph{Eigenvector centrality} measures the degree of membership of a given node in the largest core/periphery structure in the graph.\footnote{Except in very rare cases for which the graph adjacency matrix lacks a principal eigenvalue.  In such circumstances, eigenvector centrality is a signed indicator of membership in the two largest core/periphery structures (positive versus negative).} The eigenvector centrality is also the best one-dimensional approximation of the graph structure (in a least-squares sense), and accuracy in reproducing it indicates the extent to which the model is able to recover the broadest structural features of the graph. 

\item \emph{Betweenness centrality} \citep{freeman1978centrality} measures the extent to which a node lies on unique shortest paths between other nodes in the network. A node with a high betweenness centrality acts as a bridge between other parts of the network.  Betweenness is highly sensitive to the path structure in the network, and accuracy on this index thus indicates a model that is able to preserve distances among nodes.
\end{itemize}

Similar to $R^2$, the $\rho_{C}$ statistic has an upper limit of $1$, and it is equal to $1$ if and only if the proposed model reproduces networks that are perfectly correlated with the observed networks with respect to nodes' roles and positions. A smaller $\rho_{C}$ value indicates that the model is less able to recapitulate the observed network's structure. If $\rho_{C}$ is negative, then the predicted node-level indices deviate from the observed networks' values substantively even after adjusting for the variation across the observed values.

It should be noted that, like any other reliability metric, $\rho_C$ measures relative discrimination and hence is sensitive to the underlying variation in the observed centrality score.  Where there is little variation in centrality (and hence little opportunity for discrimination), it may be better to use the raw $MSE_C$ values than $\rho_C$.   The choice depends on whether the objective is to measure direct accuracy, versus ability to discriminate between high and low centrality nodes per se; in some cases, it may be worthwhile to examine both metrics.

\subsection{Graph level metrics}

A third class of selection metrics quantify how well (or poorly) the proposed model reproduces the observed networks' graph-level structural features. Specifically, we here measure the discrepancies between the observed network and the simulated networks' betweenness and degree centralization indices 
\citep{freeman1978centrality}. The centralization score is defined by
$$ C_{*}(G) = \sum_{i=1}^{n} ( max_{v' \in V} C(v') - C(i) )  $$
with $C$ being any generic centrality score. $C_{*}$ can be re-written as \break $n \left[ \max_{v' \in V} C(v') - \tfrac{1}{n} \sum_{i=1}^n C(i)\right]$, clarifying that the centrality is proportional to the difference between the maximum and mean centrality scores (i.e., the extent to which one vertex is much more central than average). In this paper, all centralization scores are normalized by the corresponding theoretical maximum centralization scores. 

For any generic graph level metric $C_{*}(G)$, we consider the root mean squared error (RMSE) as the discrepancy measure,
\begin{equation}\label{eq:RMSE_GLI}
RMSE_{C_{*}} = \sqrt[]{ \frac{ \sum_{m=1}^{M} \sum_{b=1}^{B} ( C_{*}(G^{obs}) - C_{*}(\hat{G}^{m,b})  )^{2} }{ M B} }.
\end{equation}
The RMSEs for degree and betweenness centralization then respectively indicate the extent to which the model is able to capture the upper tails of these two distributions.  Poor performance on these measures may suggest that the model lacks the ability to identify nodes that deviate substantially from the norm - a possible indicator that additional covariates are required (assuming that such identification is desired).  

\subsection{Remarks on Metrics and Held-out Strategies}
The proposed metrics cover a wide range of predictive outcomes at multiple scales, but they are not exhaustive. Our aim is to provide a suite of simple measures that are of general use. Importantly, HOPE can easily be adapted to assess the quality of models based on alternative structural features or error metrics. In fact, we encourage researchers to carefully consider their modeling objectives when deciding which metrics to use as well as which held-out data strategy to employ. Alternative metrics may provide distinct or even  contrasting information, since particular models may be better at some prediction tasks than others.  Moreover, different ways of holding out data may yield distinct results, since each both provides different information to the model during the estimation phase and poses a different prediction task during the imputation phase.  Comparing performance across tasks thus gives useful insights into a model's strengths and weaknesses.

\subsection{Efficient implementation of HOPE in parallel}
It is worth noting that the proposed HOPE procedure is highly parallel. Once the edge variables are partitioned, then each subset can be run independently. Specifically, we define the running time of HOPE as a function of the number of available computing cores $x$, number of folds $M$,  

\begin{align}
    & \text{Total running time}  =  \\ 
& \text{held-out dyad division time} + \frac{\text{ERGM fitting time} \times M}{x} + \text{result combination time}  
\end{align}

where ERGM fitting time is dependent upon the model complexity and also the number of held-out edge variables. In fact, the ERGM fitting time dominates the total run time and hence running in parallel can lead to an approximately $x$-fold reduction.  

\section{Case Studies}
\label{sec:application}
In this section, we examine the effectiveness of HOPE as a tool for model selection using two well-known data sets from the social network literature: Lazega's collaboration network between lawyers \citep{lazega1999multiplexity,lazega2001collegial} and an adolescent friendship network of $50$ girls from the ``Teenage Friends and Lifestyle Study" data set \citep{michell2000smoke}. Table \ref{tb:summary_stats_data} shows that these two networks substantially differ in their densities and degree distributions, as the collaboration network is denser and has a less skewed degree distribution, while the friendship network is more sparse and has a more skewed degree distribution. These differences produce distinct challenges for HOPE as shown below, while also demonstrating the general utility of the method. 

\begin{table}[ht]
\centering
\caption{Descriptive statsitics for Both Networks \label{tb:summary_stats_data}}
\begin{tabular}{lll}
  \hline
 & Lazega Lawyer & Teenage Friends \\ 
  \hline
Network size & 36 & 50  \\ 
  Number of edges & 115 & 74 \\ 
  Directed & No & No \\ 
  Network density & 0.18 &  0.06  \\ 
  Mean degree & 6.39 & 2.96  \\ 
  SD degree & 4.18 & 1.83  \\
  Skewness degree & 0.29 & 0.65  \\
  Number of isolates & 2 & 3  \\ 
  Transitivity & 0.39 &  0.42  \\
   \hline
\end{tabular}
\end{table}

\subsection{Collaboration Network Between Lawyers}
\begin{figure}[ht]
\centering
\includegraphics[height = 60mm, width=60mm]{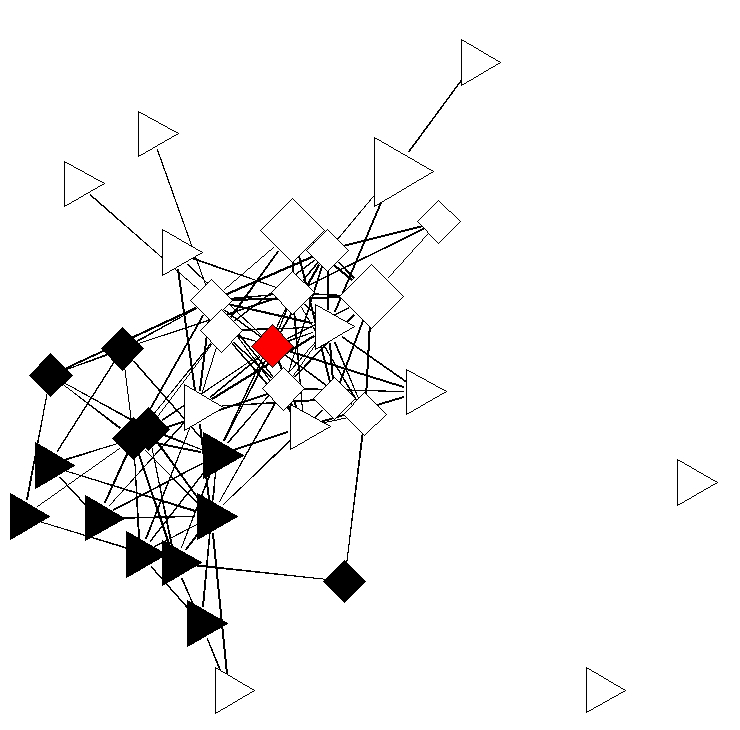}
\caption{Lazega's lawyers collaboration network. Colors indicate the different offices (White = Boston, Black = Hartford, Red = Providence); size indicates gender, men (smaller) and women (larger); shape indicates practice, ligitation (triangle), corporate (square). \label{fig:lazega_lawyer}}
\end{figure}

Lazega's lawyer data contain the collaboration relationships between $36$ partners in a New England law firm and have been used in a number of illustrative analyses \citep[e.g.,][]{snijders2006new, hunter2006inference, caimo2013bayesian}. The data set includes several nodal covariates: the \qq{Office} that the lawyer works at (out of three, 1=Boston; 2=Hartford; 3=Providence); the \qq{Practice} of the lawyer (out of two, 1=litigation; 2=corporate); the \qq{Gender} (1=Male; 2=Female) of the lawyer; and each lawyer's \qq{Seniority} within in the firm, which is their rank order of entry into the firm (lower numbers indicate more senior partners, ranging from 1 to 36). The summary statistics in Table \ref{tb:summary_stats_data} indicate that the degree distribution is relatively even and not highly skewed. Additionally, Figure \ref{fig:lazega_lawyer} illustrates a high level of homophily based on office and (to a lesser extent) practice.

We consider five competing model specifications that range from the simplest stochastic model in which all edge variables are hypothesized as i.i.d realizations of a Bernoulli random variable (i.e. homogeneous Bernoulli graph) to models that separately contain dyadic dependent terms and covariate terms. The final two models (Model 4 and Model 5) contain both types of terms, and in particular, Model 5 specifies a more granular relationship between nodes' covariates. These models are listed below as the terms used in \texttt{ergm} function (see the Appendix for more details of these network statistics) -- 
\begin{itemize}
    \item Model 1 : edges
    \item Model 2 : edges + gwesp(0.75)
    \item Model 3 : edges + nodecov(\qq{Seniority}) + nodecov(\qq{Practice}) + nodematch("Practice") + nodematch(\qq{Gender}) + nodematch(\qq{Office})
    \item Model 4 : edges + gwesp(0.75) + nodecov(\qq{Seniority}) + nodecov(\qq{Practice}) + nodematch(\qq{Practice}) + nodematch(\qq{Gender}) + nodematch(\qq{Office})
    \item Model 5 : edges + gwesp(0.75) + nodecov(\qq{Seniority}) + nodecov(\qq{Practice}) + nodematch(\qq{Practice},diff=T,keep=2) + nodematch(\qq{Gender},diff=T) + nodematch(\qq{Office},diff=T,keep=c(1,2))
\end{itemize}

The models above incrementally build upon the preceding models to generate several competing models between which HOPE can adjudicate. Model 2 builds upon model 1 by accounting for the dependence among edge variables with the inclusion of the \emph{geometrically edgewise shared partner} (GWESP) statistic with a fixed decay parameter $\phi = 0.75$, which captures the tendency for local clustering. Similarly, model 3 builds upon model 1 by including nodal covariates that capture the effects of nodes' attributes on their propensities to form ties overall and with similar types of nodes (though this is still a dyadic independent model). Model 4 contains all of the terms from the previous three models. Model 5 contains all of the terms in model 4 as well as additional terms for differential homophily for three of the nodes' attributes: \qq{Gender}, \qq{Office}, and \qq{Practice}. This final model moves towards a saturated model by replacing the uniform homophily effects with differential homophily effects. The former generates a network statistic of all nodes that are equal for a given attribute, while the latter generates a separate network statistic for each of the values for a given attribute. These models represent a range of specifications that one could consider for this network. While other models could be posited, we employ these to illustrate how HOPE can adjudicate between possible specifications.

\begin{table}[ht]
\begin{center}
\caption{Estimated Coefficients for Models Fit to the Lazega Lawyers' Collaboration Network\label{tb:mod_coef_lazega}}
\resizebox{\textwidth}{!}{
\begin{tabular}{l l l l l l }
\hline
 & Model 1 & Model 2 & Model 3 & Model 4 & Model 5 \\
\hline
Edges          &    $-1.499 \; (0.10)^{***} $     & $-3.80 \; (0.25)^{***}$ & $-8.31 \; (0.95)^{***}$ & $-7.37 \; (0.73)^{***}$ & $-4.67 \; (0.78)^{***}$ \\
GWESP, $\phi=0.75$   & & $1.05 \; (0.12)^{***}$  &                         & $0.93 \; (0.15)^{***}$  & $0.95 \; (0.16)^{***}$  \\
Main Seniority   & &                         & $0.04 \; (0.01)^{***}$  & $0.02 \; (0.01)^{***}$  & $0.02 \; (0.01)^{**}$   \\
Main Practice    & &                         & $0.90 \; (0.16)^{***}$  & $0.41 \; (0.12)^{***}$  & $-0.37 \; (0.22)$       \\
Homophily Practice  & &                         & $0.88 \; (0.23)^{***}$  & $0.76 \; (0.19)^{***}$  &                         \\
Homophily Gender    & &                         & $1.13 \; (0.35)^{**}$   & $0.70 \; (0.26)^{**}$   &                         \\
Homophily Office    & &                         & $1.65 \; (0.25)^{***}$  & $1.15 \; (0.20)^{***}$  &                         \\
Homophily Practice  (Litigation) & &                         &                         &                         & $1.50 \; (0.39)^{***}$  \\
Homophily Gender (Male)   & &                         &                         &                         & $0.48 \; (0.27)$        \\
Homophily Gender (Female)   & &                         &                         &                         & $0.59 \; (1.35)$        \\
Homophily Office (Boston)   & &                         &                         &                         & $1.00 \; (0.20)^{***}$  \\
Homophily Office (Hartford)   & &                         &                         &                         & $1.46 \; (0.24)^{***}$  \\
\hline
AIC          & 600.8        & 524.3                  & 513.8                  & 472.0                  & 470.0                  \\
BIC          & 605.2       & 533.2                  & 540.5                  & 503.2                  & 510.0                  \\
% Log Likelihood       & -260.13                 & -250.90                 & -229.02                 & -225.98                 \\
\hline
\multicolumn{6}{l}{\scriptsize{$^{***}p<0.001$, $^{**}p<0.01$, $^*p<0.05$}}
\end{tabular}}
\end{center}
\end{table}

Table \ref{tb:mod_coef_lazega} provides a summary of the estimated parameters along with the nominal AIC/BIC when each model was fit to the fully observed data. Based on the models fit to the full network, model 4 is preferred according to the nominal BIC and model 5 by the AIC. Additionally, model 3 (with node covariates) is preferred over model 2 based on the AIC, but model 2 seems to be better according to the BIC. Based on the AIC/BIC, one has conflicting information regarding which model to select for this network while knowing little about the specific strengths and weaknesses of the competing models. HOPE provides an alternate basis for model selection, which does not depend on the assumptions underlying the AIC or BIC, and allows model comparison based on their predictive performances. We note that the AIC favors the most complicated model (i.e. model 5), but model 4 has the smallest BIC compared to all of the candidate models. Such disagreement is warranted as the AIC, by design, attempts to choose a model as large as the true model, while the probability of selecting a fixed, most parsimonious, and true model tends to $1$ as the sample size tends to $\infty$ \citep[e.g.][]{nishii1984asymptotic, haughton1988choice, haughton1989size} based on the BIC. % say something about AIC/BIC

We implement HOPE on each of the models specified above, with $B=500$ as the sample size for the conditional simulations. The calculated metrics under each model specification and held-out strategy are provided in Table \ref{tb:summary_stats_lazega}. When random samples of edges are held-out, we set $M=35$ (the same number of edge variables as under node-held-out) to make the metrics comparable to those of node held-out. Throughout, we will not attempt to interpret the magnitudes of these values but focus on the relative difference and ranking between models.   
Focusing on the dyad-level metrics, the substantial differences between model 1 and model 2 suggest that accounting for the dependencies between edge variables is crucial, and the importance of incorporating nodal covariate information is indicated by the noticeable gap between the dyad-level metrics of model 1 and model 3.  Model 4, which contains both nodal covariates and dyadic dependent terms, is associated with the highest predictive accuracy of nulls (Null ACC), overall predictive accuracy (Overall ACC) and smallest total squared loss (TSL), for all of the held-out strategies. Although model 5 produces the highest predictive accuracy of edges (Edge ACC), it does not perform as well as model 4 on the other dyad-level metrics.  Overall, this presents evidence in favor of model 4, which is more parsimonious than model 5 yet still performs better on all of the other all dyad-level metrics.

If the primary objective of the proposed model is to explain phenomena that depend on variations in nodes' positions/roles, then node-level and graph-level metrics are better suited. The models are quite comparable in recapitulating nodes' degree and eigenvector centralities as well as the network's degree centralization. However, the more complex models (models 4 and 5) better predict nodes' betweenness centrality and the network's betweenness centralization index. The rankings between the models are largely consistent across the different metrics and held-out strategies. Taken together, HOPE suggests that model 4 is the preferred model (for most types of analyses) since it performs at least no worse than model 5, while having the clear advantage of being more parsimonious. 

It is worth mentioning that particular held-out strategies may be preferred depending on both modeling objectives and choices of target metric. Under leave-$1$-out, where each edge variable is held-out and assessed separately, the marginal probability of edge variables $\hat{y}_{k,l}$ can be calculated exactly via change scores \citep{snijders2006new}, allowing simple and accurate estimates of dyad-level metrics. However, because many node and graph-level metrics vary little under single edge changes, leave-1-out tends not to provide sufficient information on predictive performance with respect to these measures to discriminate between similar models.  By contrast, omitting larger numbers of edge states (either via node-held-out or leave-$M$-out strategies) poses a more difficult prediction problem, providing greater discriminatory power for comparing models' abilities to preserve higher-order structural features.  Similarly, the information provided by leave-$M$-out and node-held-out strategies differ due to the concentration of held-out edge states on specific nodes in the latter case (making node-held-out well-suited for e.g. comparing models that are intended to capture vertex-level structure).  Unsurprisingly, the calculated dyad-level metrics under leave-$M$-out are found to approximate those of leave-1-out better, as the missing data are more evenly distributed. Overall, leave-$M$-out appears to be the candidate for default choice of HOPE for general purposes, while other held-out strategies are also useful in certain scenarios.  

\newpage
\begin{sidewaystable}
\centering
\resizebox{\textwidth}{!}{\begin{threeparttable}
\caption{HOPE Metrics for 5 Models Fit to Lazega's Lawyers' Collaboration Network \label{tb:summary_stats_lazega}}
\begin{tabular}{rllll|llll|lll|ll}
  \hline
  &  & & & & & & Dyad-level & & & Node-level & & Graph-level & \\
 & Model & AIC & BIC & Type & Edge ACC & Null ACC & Overall ACC & TSL & $\rho$ Degree & $\rho$ Betweenness & $\rho$ Eigen & RMSE, Between Cen. & RMSE, Degree Cen. \\ 
  \hline
% new results are completely reproducible as I fixed the random seed for ergm fitting and simulation
%     & 1 & 600.8 & 605.2 & Leave-1-out & 0.192 & 0.817 & 0.703 & 92.55 & 0.99903 & 0.99587 & 0.99897 & 0.00496 & 0.00462 \\ 
%     & 2 & 523.0 & 531.9 & Leave-1-out & 0.388 & 0.867 & 0.779 & 69.067 & 0.99928 & 0.99793 & 0.99916 & 0.00414 & 0.00420 \\ 
%     & 3 & 513.8 & 540.5 & Leave-1-out & 0.306 & 0.845 & 0.747 & 80.701 & 0.99917 & 0.99662 & 0.99910 & 0.00456 & 0.00426 \\ 
%     & 4 & 471.8 & 502.9 & Leave-1-out & 0.43 & \textbf{0.875} & \textbf{0.794} & \textbf{66.334} & \textbf{0.99933} & 0.99805 & \textbf{0.99922} & 0.00404 & \textbf{0.00395} \\ 
%     & 5 & 470.0 & 510.0 & Leave-1-out & \textbf{0.434} & 0.873 & 0.793 & 66.54 & 0.99932 & \textbf{0.99809} & \textbf{0.99922} & \textbf{0.00403} & 0.00396 \\ 
%   \hline
    & 1 & 600.8 & 605.2 & Leave-1-out & 0.192 & 0.817 & 0.703 & 92.55 & \textbf{0.999} & 0.996 & \textbf{0.999} & 0.005 & 0.005 \\ 
    & 2 & 523.0 & 531.9 & Leave-1-out & 0.388 & 0.867 & 0.779 & 69.067 & \textbf{0.999} & \textbf{0.998} & \textbf{0.999} & \textbf{0.004} & \textbf{0.004} \\ 
    & 3 & 513.8 & 540.5 & Leave-1-out & 0.306 & 0.845 & 0.747 & 80.701 & \textbf{0.999} & 0.997 & \textbf{0.999} & 0.005 & \textbf{0.004} \\ 
    & 4 & 471.8 & 502.9 & Leave-1-out & 0.43 & \textbf{0.875} & \textbf{0.794} & \textbf{66.334} & \textbf{0.999} & \textbf{0.998} & \textbf{0.999} & \textbf{0.004} & \textbf{0.004} \\ 
    & 5 & 470.0 & 510.0 & Leave-1-out & \textbf{0.434} & 0.873 & 0.793 & 66.54 & \textbf{0.999} & \textbf{0.998} & \textbf{0.999} & \textbf{0.004} & \textbf{0.004} \\ 
   \hline
    & 1 & 600.8 & 605.2 & Leave-$M$-out & 0.18 & 0.817 & 0.701 & 94.747 & 0.963 & 0.871 & 0.962 & 0.024 & 0.026 \\ 
    & 2 & 523.0 & 531.9 & Leave-$M$-out & 0.37 & 0.863 & 0.773 & 71.45 & 0.973 & 0.932 & 0.969 & \textbf{0.02} & 0.023 \\ 
    & 3 & 513.8 & 540.5 & Leave-$M$-out & 0.307 & 0.846 & 0.748 & 80.951 & 0.968 & 0.894 & 0.966 & 0.021 & 0.023 \\ 
    & 4 & 471.8 & 502.9 & Leave-$M$-out & 0.419 & \textbf{0.872} & \textbf{0.789} & \textbf{67.412} & \textbf{0.975} & 0.935 & \textbf{0.971} & \textbf{0.02} & \textbf{0.022} \\ 
    & 5 & 470.0 & 510.0 & Leave-$M$-out & \textbf{0.422} & 0.871 & \textbf{0.789} & 67.842 & \textbf{0.975} & \textbf{0.936} & \textbf{0.971} & \textbf{0.02} & \textbf{0.022} \\ 
    \hline
    & 1 & 600.8 & 605.2 & Node & 0.179 & 0.816 & 0.7 & 94.992$^{\dag}$ & 0.943 & 0.874 & 0.938 & 0.021 & 0.021 \\ 
    & 2 & 523.0 & 531.9 & Node & 0.215 & 0.838 & 0.724 & 88.786$^{\dag}$ & 0.931 & 0.915 & 0.921 & \textbf{0.018} & 0.021 \\ 
    & 3 & 513.8 & 540.5 & Node & 0.3 & 0.841 & 0.742 & 83.656$^{\dag}$ & \textbf{0.948} & 0.885 & \textbf{0.947} & 0.019 & \textbf{0.02} \\ 
    & 4 & 471.8 & 502.9 & Node & 0.337 & \textbf{0.863} & \textbf{0.767} & \textbf{76.428}$^{\dag}$ & 0.945 & 0.92 & 0.94 & \textbf{0.018} & \textbf{0.02} \\ 
    & 5 & 470.0 & 510.0 & Node & \textbf{0.345} & 0.86 & 0.766 & 77.780$^{\dag}$ & 0.945 & \textbf{0.921} & 0.939 & \textbf{0.018} & \textbf{0.02} \\ 
    % & 1 & 600.8 & 605.2 & Node & 0.179 & 0.816 & 0.7 & 189.984$^{\dag}$ & 0.943 & 0.874 & 0.938 & 0.021 & 0.021 \\ 
    % & 2 & 523.0 & 531.9 & Node & 0.215 & 0.838 & 0.724 & 177.572$^{\dag}$ & 0.931 & 0.915 & 0.921 & \textbf{0.018} & 0.021 \\ 
    % & 3 & 513.8 & 540.5 & Node & 0.3 & 0.841 & 0.742 & 167.311$^{\dag}$ & \textbf{0.948} & 0.885 & \textbf{0.947} & 0.019 & \textbf{0.02} \\ 
    % & 4 & 471.8 & 502.9 & Node & 0.337 & \textbf{0.863} & \textbf{0.767} & \textbf{152.855}$^{\dag}$ & 0.945 & 0.92 & 0.94 & \textbf{0.018} & \textbf{0.02} \\ 
    % & 5 & 470.0 & 510.0 & Node & \textbf{0.345} & 0.86 & 0.766 & 155.56$^{\dag}$ & 0.945 & \textbf{0.921} & 0.939 & \textbf{0.018} & \textbf{0.02} \\ 
    \hline
\end{tabular}
% }
    \begin{tablenotes}
    \scriptsize{
    \item[$\dag$] The original TSL is scaled by $2$ for the purpose of comparison, as each edge variable is held-out twice in the entire process under node held-out.}  
    \end{tablenotes}
\end{threeparttable}} %
\end{sidewaystable}

\subsection{Teenage Friendship Network}
\begin{figure}[H]
\centering
\includegraphics{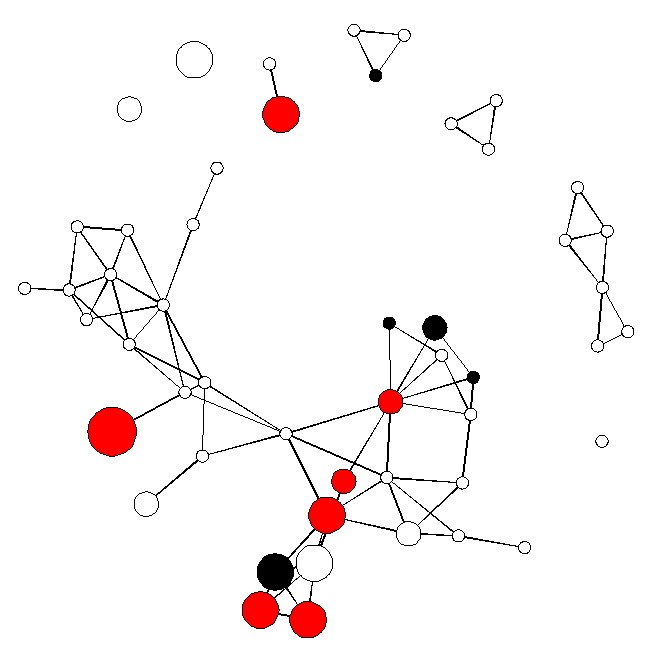}
\caption{Teenage friendship network. Colors indicate individuals' different smoking patterns (white = non-smoker; black = occasional smoker; and red = regular smoker); size indicates individuals' frequencies of drug use (larger nodes indicate greater frequencies). \label{fig:teenage}}
\end{figure}

The teenage friendship network contains undirected friendship ties between $50$ young women\footnote{Following \citep{bouranis2018bayesian}, we use the intersection of reported ties and symmetrize the adjacency matrix.}. This network is a subset of the \qq{Teenage Friends and Lifestyle Study} data set \citep{michell2000smoke}, which has three years of friendships from 1995 to 1997. We use the data from the first year along with two nodal attributes: \texttt{smoke}, which takes the values 1 (non-smoker), 2 (occasional smoker), and 3 (regular smoker); and \texttt{drugs} for cannabis use, which takes the values 1 (non-drug user), 2 (tried once), 3 (occasional) and 4 (regular use).  Prior analysis on these data \citep{bouranis2018bayesian} dichotomized the \texttt{drugs} such that 1 indicates individuals that never used drugs or tried once; and 2 indicates occasional or regular drug use.  We only use the dichotomized version of the variable, which we refer to as \texttt{drugs\_binary}. Figure \ref{fig:teenage} depicts the friendship network and individuals' frequencies of smoking and drug use. As shown, most friendships are homophilous based on smoking and drug use, with one individual bridging two distinct groups of the largest component.
%  The network has one large component comprised of 33 individuals along with several smaller components and a few isolates.
%The adolescent friendship network is a subgraph of a friendship network among 50 girls collected in the ``Teenage Friends and Lifestyle Study" data set \citep{michell2000smoke} available at \url{https://www.stats.ox.ac.uk/~snijders/siena/Glasgow_data.htm}. The original data set records the friendships over three years from 1995 to 1997. In this study, we only consider the friendship network in 1995 (the first time point) and two node-level attributes: \texttt{smoke}, which takes values 1 (non-smoker), 2 (occasional smoker), 3 (regular smoker) and \texttt{drugs} (Cannabis use), which takes values 1 (non-drug user), 2 (tried once), 3 (occasional) and 4 (regular drug user). The network was used as a case study in \citet{bouranis2018bayesian} to demonstrate Bayesian model selection of ERGMs via Adjusted Pseudolikelihoods, where the covariate \texttt{drugs} was dichotomized as a binary variable (here we name it as \texttt{drugs\_binary}), with 1 denoting non-drug user or tried once and 2 representing occasional or regular drug user. Figure \ref{fig:teenage} shows the friendship network according to the attributes drug usage and smoking status. We can see some homophily in friendships by smoking and drug usage behavior as nodes of the same color and same size appear to have a higher tendency to form links.

Again, we evaluate five models that incrementally build upon the previous models, similar to the models considered for the lawyers' collaboration network.  Model 5 includes a differential homophily term for smoking that is absent from Models 3 and 4. The decay parameters of GWESP and GWDEG terms are fixed as $\log(2)$ and $0.8$ following the models used in \citet{bouranis2018bayesian}.

\begin{itemize}
    \item Model 1: edges
    \item Model 2: edges + gwesp(log(2)) + gwdegree(0.8)
    \item Model 3: edges + nodematch( \qq{drugs\_binary}, diff=T )
    % \item Model 4: edges + gwesp(log(2)) + gwdegree(0.8) + nodematch( \qq{drugs\_binary}, diff=F )
    \item Model 4: edges + gwesp(log(2)) + gwdegree(0.8) + nodematch( \qq{drugs\_binary}, diff=T ) % mod 22
    \item Model 5: edges + gwesp(log(2)) + gwdegree(0.8) + nodematch( \qq{drugs\_binary}, diff=T ) + nodematch(\qq{smoke}, diff=T)
\end{itemize}

\begin{table}[ht]
\begin{center}
\caption{Estimated Coefficients for Models Fit to the Teenage Friendship Network \label{tb:mod_coef_teenage}}
\resizebox{\textwidth}{!}{
\begin{tabular}{l l l l l l}
\hline
 & Model 1 & Model 2  & Model 3 & Model 4 & Model 5 \\
\hline
Edges           &            $-2.74 \; (0.12)^{***}$ & $-6.12 \; (0.43)^{***} $ & $-4.01 \; (0.41)^{***}$ & $-6.72 \; (0.55)^{***}$ & $-6.75 \; (0.54)^{***}$ \\
Homophily drug (Never/Tried Once) &   &  & $1.42 \; (0.43)^{**}$  & $0.82 \; (0.35)^{*}$    & $0.75 \; (0.36)^{*}$    \\
Homophily drug (Occasional/Regular Use)  & &  & $3.09 \; (0.59)^{***}$  & $1.57 \; (0.35)^{***}$  & $1.81 \; (0.40)^{***}$  \\
GWESP, $\phi = \log(2)$ &                       & $1.783 \; (0.18)^{***}$ &   & $1.71 \; (0.19)^{***}$  & $1.71 \; (0.18)^{***}$  \\
GWDEG, $\phi = 0.8$ &                                      & $2.246 \; (0.43)^{***}$ &                       & $2.20 \; (0.45)^{***}$  & $2.21 \; (0.45)^{***}$  \\
Homophily smoke (Nonsmoker) & &                                    &                         &                         & $0.13 \; (0.25)$        \\
Homophily smoke (Occasional Smoker)  & &                                  &                         &                         & $0.93 \; (0.75)$        \\
Homophily smoke (Regular Smoker)  &  &                                 &                           &                         & $-0.69 \; (0.90)$       \\
\hline
AIC & 560.8  & 453.6                            & 535.1                  & 438.7                  & 443.7                  \\
BIC & 565.9  & 469.0                           & 550.4                  & 464.2                  & 484.6                  \\
% Log Likelihood                & -264.53                 & -214.34                 & -213.84                 \\
\hline
\multicolumn{6}{l}{\scriptsize{$^{***}p<0.001$, $^{**}p<0.01$, $^*p<0.05$}}
\end{tabular}}
\end{center}
\end{table}

Table \ref{tb:mod_coef_teenage} provides a summary of the estimated parameters along with the nominal AIC/BIC when each model was fit to the fully observed data. Based on the models fit to the full network, model 4 is preferred according to the nominal AIC/BIC. For each model specified above, we implement HOPE with $B=500$ as the sample size for the conditional simulations. Table \ref{tb:summary_stats_teenage} displays the calculated metrics for each model using the different held-out strategies, along with the AIC and BIC when fit to the original network. When random samples of edges are held-out, we set $M=49$ (the same number of edge variables as under node-held-out) to make the metrics comparable to those of node held-out. In contrast to the lawyer network, the inclusion of the dyadic dependent term improves the fit of the models to the friendship network significantly more than including nodal covariates according to the AIC/BIC. 

The dyad-level metrics unanimously indicate that model 4 is as good if not better than all other models. It is worth noting that model 2 is itself a compelling model as the gaps between the dyad-level metrics for model 2 and model 4 are minimal, suggesting that dyadic dependent terms play a more central role in explaining the formation of this friendship network. Similar to the collaboration network, the most complex model (model 5) does not lead to a better predictive performance at the dyad level. 

Larger variations on the node-level and graph-level metrics are present in this case. This is partly due to the sparsity of this friendship network, which makes the misplacement of a single edge generate a more substantial effect on the structural features. Such variation can be more informative for the purpose of model comparison. We find that all of the models, including the dyadic independence models (models 1 and 3), are fairly effective in reproducing nodes' degree and eigenvector centralities. However, these models perform less well at recovering measures such as betweenness, compared to other models with terms that account for dyadic dependence.  This may be because the combination of sparsity and local clustering that characterizes the network (and to which betweenness is sensitive) cannot be easily captured without GWESP.

The differences between the results of this network and the Lazega network are revealing. Note that the improvement produced by the covariate effects alone is not as substantial as that by dependence terms compared to the baseline homogeneous Bernoulli graph. In fact, the predictive performance of model 2 is almost comparable to those of models with both covariates and dyadic dependent terms under leave-$M$-out sampling. This suggests a dominant role of endogenous network effects in the formation of the teenage friendship network, as opposed to the strong role for covariates in structuring collaboration among lawyers. In addition, the substantially different performances of model 2 under node and leave-$M$-out held-out strategies further suggests the importance of dyadic dependence in network structure, as local structure about each vertex appears to be largely emergent and hence difficult to predict without access to other edge variables on the same vertex.  By contrast, if tie formation were dominated by observed covariate effects, we would see much less differentiation between leave-$M$-out and node-held-out performance.  Such performance comparisons thus provide a useful tool for inferring the relative importance of different mechanisms in determining network structure.

% \begin{table}[ht]
\begin{sidewaystable}
\centering
\resizebox{\textwidth}{!}{\begin{threeparttable}
\caption{HOPE Metrics for 5 Models Fit to Teenage Friendship Network \label{tb:summary_stats_teenage}}
\begin{tabular}{rllll|llll|lll|ll}
% {rlllllllllllll}
  \hline
    &  & & & & & & Dyad-level & & & Node-level & & Graph-level & \\
 & Model & AIC & BIC & Type & Edge ACC & Null ACC & Overall ACC & TSL & $\rho$ Degree & $\rho$ Betweenness & $\rho$ Eigen & RMSE, Between Cen. & RMSE, Degree Cen. \\ 
  \hline
   & 1 & 560.8 & 565.9 & Leave-1-Out & 0.072 & 0.930 & 0.878 & 69.30 & 0.999 & 0.986 & 0.997 & 0.011 & 0.002 \\
   & 2 & 453.6 & 469.0 & Leave-1-Out & 0.480 & 0.965 & 0.936 & 41.064 & 0.999 & 0.998 & 0.998 & \textbf{0.004} & \textbf{0.001} \\
   & 3 & 535.1 & 550.4 & Leave-1-Out & 0.088 & 0.948 & 0.896 & 66.477 & 0.999 & 0.989 & 0.997 & 0.010 & 0.002 \\
   & 4 & 438.7 & 464.2 & Leave-1-Out & \textbf{0.483} & \textbf{0.968} & \textbf{0.938} & \textbf{39.982} & \textbf{0.999} & \textbf{0.998} & 0.998 & \textbf{0.004} & \textbf{0.001} \\
   & 5 & 443.7 & 484.6 & Leave-1-Out & 0.480 & 0.966 & 0.937 & 40.684 & 0.999 & 0.998 & \textbf{0.998} & \textbf{0.004} & \textbf{0.001} \\ 
   \hline
   
   & 1 & 560.8 & 565.9 & Leave-$M$-Out & 0.062 & 0.939 & 0.886 & 69.528 & 0.932 & 0.355 & 0.895 & 0.064 & 0.013 \\ 
   & 2 & 453.6 & 469.0 & Leave-$M$-Out & 0.457 & 0.962 & 0.932 & 43.365 & 0.958 & \textbf{0.892} & 0.906 & \textbf{0.027} & \textbf{0.009} \\ 
   & 3 & 535.1 & 550.4 & Leave-$M$-Out & 0.086 & 0.941 & 0.889 & 68.007 & 0.934 & 0.35 & 0.896 & 0.065 & 0.012 \\ 
   & 4 & 438.7 & 464.2 & Leave-$M$-Out & \textbf{0.463} & \textbf{0.964} & \textbf{0.933} & \textbf{42.664} & \textbf{0.959} & 0.884 & 0.909 & 0.028 & \textbf{0.009} \\ 
   & 5 & 443.7 & 484.6 & Leave-$M$-Out & \textbf{0.463} & 0.963 & \textbf{0.933} & 42.897 & \textbf{0.959} & 0.884 & \textbf{0.911} & \textbf{0.027} & \textbf{0.009} \\ 
   \hline
   
   & 1 & 560.8 & 565.9 & Node & 0.058 & 0.94 & 0.886 & 140.246$^{\dag}$ & \textbf{0.927} & 0.477 & \textbf{0.857} & 0.056 & \textbf{0.01} \\ 
   & 2 & 453.6 & 469.0 & Node & 0.076 & 0.94 & 0.887 & 136.863$^{\dag}$ & 0.923 & \textbf{0.864} & 0.766 & \textbf{0.035} & 0.011 \\ 
   & 3 & 535.1 & 550.4 & Node & 0.081 & 0.941 & 0.889 & 137.486$^{\dag}$ & \textbf{0.927} & 0.483 & 0.856 & 0.055 & \textbf{0.01} \\ 
   & 4 & 438.7 & 464.2 & Node & \textbf{0.101} & \textbf{0.943} & \textbf{0.892} & \textbf{132.847}$^{\dag}$ & 0.925 & 0.859 & 0.779 & \textbf{0.035} & 0.011 \\ 
   & 5 & 443.7 & 484.6 & Node & 0.097 & \textbf{0.943} & \textbf{0.892} & 135.706$^{\dag}$ & 0.922 & 0.859 & 0.778 & \textbf{0.035} & 0.011 \\
   \hline
   \hline
\end{tabular}
\begin{tablenotes}
    \scriptsize{
    \item[$\dag$] The original TSL is scaled by $2$ for the purpose of comparison, as each edge variable is held-out twice in the entire process under node held-out.
    }
    \end{tablenotes}
    \end{threeparttable}}
\end{sidewaystable}
%%%%%%%%%%%%%%%%%%%%%%%%%%%%%%%%%%%%%%%%%%%%%%%%%%%%%%%%%%%%%%%%%%
\section{Discussion and Conclusion}
In this paper, we employ a technique analogous to cross-validation for models with dependent, relational data: Held-Out Predictive Evaluation (HOPE). Researchers can use HOPE to compare different model specifications of ERGMs and then select the best model for their purpose. HOPE is flexible to multiple ways of holding-out data, and allows models to be evaluated based on an array of predictive metrics. Those we have delineated are by no means exhaustive, and we hope that researchers will consider other metrics in the future based on their specific research questions, such as goodness-of-fit metrics related to a graph's spectrum by \citet{shore2015spectral} if e.g. macroscopic community structure is of substantial interest.  Taken together, HOPE complements existing model adequacy checks while addressing a current limitation for model selection. Unlike other model selection tools (e.g.; AIC or the Bayes factor), HOPE is always straightforward to implement, and can be easily extended to ERGMs in which edges are counts, ranks, or continuous \citep[see e.g.,][respectively]{krivitsky2012exponential,krivitsky2017exponential, desmarais2012statistical} rather than just binary relations.  Researchers can also use HOPE to diagnose different models' predictive abilities. This allows model selection to proceed based on tasks that are more appropriate to the analyst's modeling objectives.
 
The application of HOPE to two different networks with vastly different structures indicates the importance of properly choosing the held-out strategy and metrics that will be used to select the model. Based on empirical findings in this paper, the leave-1-out strategy has the potential to yield the most reliable estimates of the dyad-level metrics, but can generate node-level and graph-level metrics that have too little variation across models, as some structural measures are fairly robust and hence remain unchanged given the misplacement of a single edge variable. Holding out each node's edge variables can alter the network structure to a large extent, and hence facilitate a more fruitful comparison of node-level and graph-level metrics, but the information it delivers regarding the predictive accuracy of edges can be drastically different from the other held-out strategies for networks that are sparse or contain imbalanced covariates. The leave-$M$-out strategy provides a better general purpose strategy, since more held-out edge variables are present in each batch than leave-1-out, and they are more evenly distributed rather than all of the dyads containing on of the vertices, which can make the assessments dependent on nodes' idiosyncracies. As a general starting point, we suggest setting $M$ equal to $n-1$ (or, in the directed case, $2n-2$), to facilitate comparison with node-held-out performance.

As noted, HOPE is distinct from other model selection and adequacy techniques.  Simulation-based goodness-of-fit approaches \citep{hunter2008goodness} are widely used for adequacy checking, but are not well-suited to model selection.  The method developed by \citet{koskinen2018outliers} can successfully identify outliers that influence coefficients' magnitude and significance, but it focuses on quantifying how outliers affect parameter estimates for a model rather than a model's predictive performance.  The AIC and BIC scores \emph{are} intended for model selection, but are difficult or impossible to calculate accurately in the case of dependence models (with only nominal data degrees of freedom currently known) and draw upon asymptotic arguments of dubious validity in typical ERGM applications. Bayes factors have better theoretical properties, but are even more difficult to compute.  All of these methods have assets, and we do not argue against their use. However, HOPE offers distinct gains in feasibility and interpretability, and it has the advantage of focusing attention on models' out-of-sample predictive abilities. As such, it moves away from coefficient tests by illuminating which effects are \emph{predictive} rather than \emph{significant}. The models for the collaboration network show that several of the heterogeneous homophily terms in model 5 are significant, and when fit to the full network, this model has a better AIC value than model 4, which only includes uniform homophilly terms.  Yet, the predictive metrics illuminate the fact that these effects are ephemeral, and more importantly, they do not improve the model's predictive accuracy compared to model 4.  HOPE's ability to illuminate these differences speaks to the ongoing controversy about significance and $p$-values \citep{goodman2019getting, kuffner2019p, mcshane2019abandon}.  In our view, terms in a model should typically improve the ability of the model to predict the phenomena of interest rather than simply being statistically significant and, by turn, significant effects that add nothing to prediction should be viewed with suspicion.  

In sum, HOPE is complementary to extant methods of model selection and model adequacy checks. We believe that the strengths of HOPE will prove to be a useful tool for research when selecting between multiple model specification of ERGMs or other statistical network models with complex dependencies.  

\section*{Acknowledgments}
This work was supported in part by NSF awards IIS-1526736 and DMS-1361425. 

\section*{Appendix A}
\label{sec:appendix}
Here, we list the definitions of the sufficient statistics used in Section \ref{sec:application}.  All of the terms are in the context of an undirected network, since that is the case for all of the data sets used. For more details, see \citet{morris2008specification} for a more comprehensive list of sufficient statistics for ERGMs. \newline

\emph{Edge statistic} (\texttt{Edges}) \newline

\begin{equation*}
\label{eq:ergm_edges}
S(y) = \sum_{i<j} y_{ij}
\end{equation*}

\emph{Geometrically weighted edgewise shared partner} (\texttt{GWESP}) statistic \newline
\begin{equation*}
\label{eq:ergm_gwesp}
w(y, \phi) = e^{\phi} \sum_{k=1}^{n-2} \left\{ 1 - (1-e^{-\phi})^{k}   \right\} EP_{k}(y)
\end{equation*}

where $EP_{k}(y)$ is the number of connected pairs that have exactly $k$ common neighbors, which is a measure of local clustering in a network. The decay parameter $\phi$ controls the relative contribution of $EP_{k}(y)$ to the GWESP statistic. \newline

\emph{Geometrically weighted degree statistic} (\texttt{GWDEG}) \newline
\begin{equation*}
\label{eq:ergm_gwesp}
v(y, \phi) = e^{\phi} \sum_{k=0}^{n-1} \left\{ 1 - (1-e^{-\phi})^{k}   \right\} D_{k}(y)
\end{equation*}

where $D_{k}(y)$ is the number of nodes who have exactly $k$ neighbors. The decay parameter $\phi$ controls the relative contribution of  $D_{k}(y)$ to the GWDEG statistic. \newline

\emph{Uniform homophily effect} (\texttt{nodematch}) \newline

Categorical covariate $x$ is defined for each node in the graph.  The term counts the total number of edges between nodes with the same values of $x$,

\begin{equation*}
\label{eq:ergm_nodematch_sex}
H(y;x) = \sum_{i < j} y_{ij}\mathbbm{1}_{ \{x_{i} = x_{j}\} } 
\end{equation*} \newline

\emph{Differential homophily effect} (\texttt{nodematch}) \newline

Categorical covariate $x$ is defined for each node in the graph, which can take $K$ distinct values, for any given possible value of $x=k$ where $k =1,2,\cdots, K$

\begin{equation*}
\label{eq:ergm_nodematch_sex}
H(y;x,k) = \sum_{i < j} y_{ij}\mathbbm{1}_{ \{x_{i} = x_{j} = k\} } 
\end{equation*} \newline

\emph{Edge-level covariate effect} (\texttt{edgecov}) \newline

Covariate $x$ is measured for each pair of nodes in the graph,

\begin{equation*}
\label{eq:ergm_edgecov}
EC(y;x) = \sum_{i<j} y_{ij}x_{ij}
\end{equation*}

\emph{Node covariate} (\texttt{nodecov}) \newline

Covariate $x$ is measured for every node in the graph, which can take numerical values

\begin{equation*}
\label{eq:ergm_edgecov}
NC(y;x) = \sum_{i<j} y_{ij} ( x_{i} + x_{j} )
\end{equation*}

%\section*{References}
\bibliographystyle{elsart-num}
\bibliography{mybibfile}
\end{document}